# Resource allocation for D2D-Based AMI Communications Underlaying LTE Cellular Networks

H. H. Esmat*, Mahmoud M. Elmesalawy, I. I. Ibrahim

Dept. of Electronics and Communications, Faculty of Engineering, Helwan University Cairo, Egypt
*eng.haisam2009@h-eng.helwan.edu.eg

**Abstract:** Smart meters are utilized to transmit the consumption information to the metering data management system for observing and management in smart grid advanced metering infrastructure systems. In the meantime, for efficient utilization for spectrum, Device-to-Device (D2D) communications underlaying LTE networks are a promising wireless communication technology for advanced metering infrastructure which supporting a technique for reusing the same radio resources (RRs) of LTE networks. Therefore, we examine the utilization of D2D communication technology for advanced metering infrastructure communications underlaying LTE networks. A novel approach is suggested for provisioning the mandatory communication between serving data concentrator and its set of smart meters using this technology. The suggested approach is dependent on two main stages. The group of permissible cellular user equipment reuse candidates for every smart meter is calculated with taking the quality of service demands for cellular user devices and smart meters into consideration in the first stage. The optimal RR allocation for every smart meter is determined based on maximizing the access rate of smart meters which can be accepted and operated in D2D reuse mode in the second stage. Simulation results prove the efficacy of the suggested approach for efficient advanced metering infrastructure communication underlaying LTE systems with accepting remarkable number of SMs and accomplishing outstanding throughput gain.

**Keywords**-Smart Grid; D2D; LTE Cellular systems; AMI; Resource Allocation; Advanced Metering Infrastructure

## 1. Introduction

Advanced metering infrastructure is considered part of the crucial parts in contemporary smart grids that are becoming an essential for the development of the recent power grids. The Advanced metering infrastructure systems are mainly based on automatic meter reading to gather, analyze, and determine energy consumption [1]. The real-time information feed backed from smart meters (SMs) to the metering data management system (MDMS) lead to significant benefits, including accurate billing and lessening of the peak power request by executing request response management programs [2]. This dealing with SMs requires to be implemented via some communication technology.

The widespread communications technologies, which can be utilized to apply advanced metering infrastructure (AMI) communications using diverse transmission media such as cellular, WiMAX, ZigBee, microwave, Wi-Fi, and power line Communication (PLC) [3-8]. These technologies include up-to-date wired and wireless communication network technologies. PLC offers a cost-effective solution which enables broadcast of smart meter (SM) information through the current power lines. However, the main constraint of PLC is its lessened data rates and the considerable alteration of the power line channel features as a result of diverse wiring practices and loads linked to the network [9]. Consequently, wireless communications are broadly viewed as a marginal solution for advanced metering infrastructure [10]. Advanced cellular networks such as long term evolution (LTE) appears as a promising technology for wide-area AMI communications. Their ubiquitous coverage permits smart metering deployments to extent over vast regions and be connected into the same management system [11], [12]. In addition, the characteristics of LTE networks in terms of supported data rate, high system reliability and low latency, facilitate critical applications within the AMI systems. The Physical Resource Block (PRB) in LTE networks is considered as the smallest radio resource element, which can be assigned to the end terminal that is specified in the 3rd Generation Partnership Project (3GPP) documents. Since SMs are frequently send small amount of information at a specified period, the assignment of a whole PRB to each SM will result in degeneration in the spectrum efficiency. Therefore, using two-tier approach for AMI communications is suggested in which the data for each group of smart meters is collected by one data concentrator (DC) through the use of one of the short-range communication technologies [13]. After that, these DCs relay the gathered data to the MDMS through LTE uplink (UL) connection. Using DCs will permit various SMs to simultaneously reuse the same PRB and thus increase the spectral efficiency.

Numerous short-range wireless communication technologies (for instance, Wi-Fi and Zigebee) can be used in the first tier for providing the communication between each DC and its associated SMs. Since these technologies using unlicensed spectrum, the interference management will become a dilemma [14]. Otherwise, D2D communication technology is considered as a promising solution for short-distance wireless communication underlaying LTE networks to improve the spectrum utilization and thus escalation the system throughput [15-18]. The notion of D2D communication is dependent on permitting the closeness devices to create a direct connection under the control of eNodeB (eNB).

The devices in LTE networks empowered by D2D communication technology can be worked in three modes. These modes are cellular, dedicated and reuse mode [19]. The cellular mode permits the ordinary method of sending information through eNB, whilst a portion of the existing resources is assigned to D2D communications in the dedicated mode. The interference management techniques used in cellular and dedicated modes are very easy and candid, but they do not achieve a remarkable efficient utilization of the presented spectrum [20]. Otherwise, the physical resource blocks assigned to CUEs are permitted to be shared by D2D devices in the reuse mode. Even though this mode enhances the utilization of the overall system spectrum, the interference among cellular transmissions and D2D transmissions might be presented. However, this problem can be controlled as the whole system is yet managed through the core of cellular system. The interference problem in reuse mode has been widely investigated in existing works [17], [21-28].

Unlike other short-range communication technologies, D2D communications underlaying LTE operates in licensed spectrum and the radio resources are suitably controlled by the system, to diminish the interference and maximize the whole network performance. Moreover, using D2D communication for providing the connectivity among every set of SMs and its serving DC will lead to numerous advantages in comparison with other stated technologies. This include automated secure authentication during the connection, automatic pairing between SMs and DC which controlled by eNB and finally the longer communication distances and higher data rates supported by D2D communication compared to the other mentioned technologies.

D2D communication technology is adopted to provision the essential communication among serving DC and its each group of SMs underlaying the cellular networks in this paper. An UL radio resource allocation (RRA) technique with quality of service providing is suggested in order to support the AMI communications underlaying the cellular networks. The suggested algorithm is dependent on two key phases. The group of permissible cellular reuse candidates for every smart meter is calculated taking the quality of service demands for smart meters and CUEs into consideration in the first stage. The optimal RRA for every smart meter is determined with the purpose of maximizing the overall number of accepted SMs in the second stage.

The paper is structured as follows. The network architecture and the problem formulation are presented in section II. Next, the suggested RRA scheme for SMs is introduced. After that, the simulation analysis and discussion are provided. Eventually, the paper in concluded in section V.

## 2. System architecture and problem formulation

Firstly, the system architecture is presented and afterwards the RRA problem for SMs is formulated in this section.

### 2.1. System Architecture

In this subsection, we introduce the suggested network architecture for AMI connectivity using D2D communications technology underlaying cellular networks. The architecture is dependent on reusing the UL RRs of cellular system for providing the mandatory connectivity among every set of SMs and its serving DC. In this paper, a cell containing a number of CUEs and several DCs serving a number of SMs is deemed. Every SM belonging to a specific set will transmit his information to its serving DC by reusing the UL RR allocated for CUEs in the cell. Each DC gathers the information from all SMs in his set and relays it to the eNB utilizing the available UL RRs.

Let $C = \{c^m | m = 1, 2, 3, ..., M\}$ characterizes the set of CUEs in the cell, $D = \{d^l | l = 1, 2, 3, ..., L\}$ represents the set of DCs and $S = \{s^n | n = 1, 2, 3, ..., N\}$ describes the set of SMs per DC, where $M$, $L$ and $N$ are the cardinalities of $C$, $D$ and $S$ respectively. Each DC is assumed to serve an equal number of SMs ($N$). The spectrum of the LTE network is divided into a number of channels with the same size called PRBs. Every CUE is assigned one distinct RB in the UL duration with no interference occurs between CUEs in our architecture. The SMs associated to one DC have the same destination which is its serving DC. Therefore, so as to shun the harmful interference among SMs communications associating to the same group at the DC, the RRA procedure ought to be controlled in such a manner to avert reusing of the same PRBs by more than one SM at the same time. On the other hand, smart meters associated to different DCs can share the same radio resources under quality of service constraints. The transmit power of CUEs and SMs are assumed to be constants, $P_C$ for CUEs and $P_S$ for SMs. We define $\Gamma_S = \{\Gamma_s^n | n = 1, 2, 3, ..., N\}$ and $\Gamma_C = \{\Gamma_c^m | m = 1, 2, 3, ..., M\}$ as the set of minimum signal-to-interference plus noise ratio (SINR), which have to be achieved to satisfy the quality of service demands of SMs and CUEs, respectively. The eNB controls the CUEs that are presumed distributed uniformly in the cell. In Fig. 1 the system architecture for AMI connectivity by using D2D communication underlaying the cellular system is illustrated, where black links exemplify the communication links for SMs and CUEs. Otherwise, red lines exemplify the interference from both SMs on eNB and from CUEs on DCs as well as interference among SMs due to resource sharing.

Because of sharing the same UL RRs of CUEs to provision the communication between SMs and DCs, the transmission from CUEs to eNB leads to interference on the transmission of SMs at the DC. Simultaneously, the transmissions from SMs to specific DC act as interferers on CUEs transmission at eNB and also act as interferers on SMs transmission that sharing the same radio resources on other DCs. Consequently, appropriate assignment for UL RRs ought to be deemed by eNB for handling the interference among CUEs and SMs and attaining the mandatory quality of service requirements for CUEs and SMs.

### 2.2. Problem Formulation

A SM can communicate with its serving DC by sharing the same UL RRs of a specific CUE once the quality of service requirements for CUE and SM can be assured. In this situation, this SM is called as an acceptable SM and the CUE that its RRs are reused is called a reuse candidate. We describe matrix $R_{(NL \times M)} = \underset{l=1}{\overset{L}{\|}} R^l$, where $R^l = [r_{n,m}^l]$ as the permissible reusing matrix that shows the set of permissible reuse candidates for every $SM\#n$ in each $DC\#l$ that satisfies the required quality of service constraints.

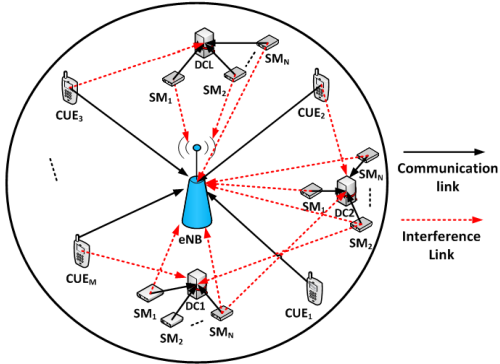

*Fig. 1. System architecture for AMI connectivity using D2D communications underlaying LTE networks.*

For each $DC\#l$, if $CUE\#m$ is a reuse candidate for $SM\#n$, then $r_{n,m}^l = 1$; otherwise, $r_{n,m}^l = 0$. Since every CUE can be permitted to be a reuse candidate for more than one SM, optimal assignment of RRs for SMs is obligatory. We describe matrix $A_{(NL \times M)} = \underset{l=1}{\overset{L}{\|}} A^l$, where $A^l = [a_{n,m}^l]$ as the resource allocation matrix as,

$$a_{n,m}^l = \begin{cases} 1, & \text{if resources of CUE } m \text{ is allocated} \\ & \text{to SMn serving by DCl} \\ 0, & \text{otherwise} \end{cases} \quad (1)$$

Because the key aim is to permit as many as SMs to be accepted and operated in the D2D reuse mode with a assured quality of service constraints, the optimal RRA problem can be mathematically described as

$$A^* = \underset{A}{\arg\max} \sum_{m=1}^{M} \sum_{l=1}^{L} \sum_{n=1}^{N} r_{n,m}^l \cdot a_{n,m}^l \quad (2)$$

S. t.,

$$\sum_{n=1}^{N} a_{n,m}^l \leq 1, a_{n,m}^l \in \{0,1\}, 1 \leq l \leq L, 1 \leq m \leq M \quad (2a)$$

$$\sum_{m \in M} a_{n,m}^l \leq 1, a_{n,m}^l \in \{0,1\}, 1 \leq n \leq N, 1 \leq l \leq L \quad (2b)$$

$$SINR_n^l \geq \Gamma_s, \quad 1 \leq l \leq L, 1 \leq n \leq N. \quad (2c)$$

$$SINR_{eNB}^m \geq \Gamma_c^m, \quad 1 \leq m \leq M. \quad (2d)$$

Constrains in (2a) ensure that no sharing between SMs that are associated to the same DC. (2b) is imposed to guarantee that every SM can share the RRs of at most one CUE. Constraints in (2c) and (2d) are quality of service constraints for SMs and CUEs, respectively.

## 3. Radio Resource Allocation (RRA) for SMs

The suggested RRA algorithm is dependent on two key phases. The various reuse candidates for every SM under the quality of service demands for SMs and CUEs are determined in the first phase. The optimal RRA for every SM is calculated according to the resulted reuse candidates for every SM with the aim of maximizing the total number of SMs that can be accepted to operate in the D2D reuse mode in the second phase. The next parts describe the two phases (stages).

### 3.1. Phase1: Interference Management and SMs Reuse Candidates

In the suggested network, there are three kinds of interference because of reusing the UL RRs of CUEs by SMs. The received signals of CUEs at eNB as a result of the SMs communications is the first type of the interferences. The second one is the interference presented on the received signals of SMs at the serving DC because of CUEs communications. The last one is the interference among various SMs that sharing the same radio resources. These three sorts ought to be controlled precisely to assure the compulsory quality of service demands for SMs and CUEs.

To shun any detrimental interference from SMs on the cellular communications as a result of reusing the same resources, a maximum interference boundary is decided for each CUE. This boundary can be determined in accordance with the minimum permissible SINR received at eNB for each $CUE\#m$ ($\Gamma_c^m$). Because all CUEs are supposed to communicate by the same power value $P_c$, the channels ought to be optimally assigned for SMs to attain the constraint in (3).

$$\frac{P_c \cdot h_{m,eNB}^m}{W^m + \sum_{l=1}^{L} \sum_{n=1}^{N} a_{n,m}^l \cdot P_s \cdot h_{n,eNB}^{l,m}} \geq \Gamma_c^m, 1 \leq m \leq M, \quad (3)$$

where, $W^m$ is the thermal noise power on $channel\#m$, $h_{m,eNB}^m$ is the channel gain from $CUE\#m$ to the eNB on $channel\# m$ and $h_{n,eNB}^{l,m}$ is the channel gain from $SM\#n$ which serving by $DC\#l$ to the eNB on $channel\# m$. Consequently, provided the mandatory SINR requirement for each $CUE\#m$ and its transmitting power, the maximum permissible interference level on $CUE\#m$ at $eNB$ ($I_{max,eNB}^m$) can be computed as

$$I_{max,eNB}^m = \frac{P_c \cdot h_{m,eNB}^m}{\Gamma_c^m} - W^m, \quad 1 \leq m \leq M \quad (4)$$

The calculated value for $I_{max,eNB}^m$ is utilized in the suggested RRA approach as a constraint for the maximum allowable interference on each $CUE\#m$. We also assume that each SM has minimum acceptable SINR $\Gamma_s$ which is equal for all SMs. Therefore, to avert any detrimental interference on SMs and ensure the mandatory quality of service requirement for each SM, the SINR constraint in (5) should be fulfilled.

$$\frac{P_s \cdot h_{n,l}^m}{P_c \cdot h_{m,l}^m + \sum_{\substack{k=1 \\ k \neq l}}^{L} \sum_{x=1}^{N} a_{x,m}^k \cdot P_s \cdot h_{x,l}^{k,m} + W^m} \geq \Gamma_s, \quad (5)$$

$$1 \leq l \leq L, 1 \leq n \leq N$$

where, $h_{n,l}^m$ is the channel gain from $SM\#n$ to its serving $DC\#l$ on $channel\#\,m$, $h_{m,l}^m$ is the channel gain for interference link from $CUE\#m$ to the $DC\#l$ on $channel\#\,m$ and $h_{x,l}^{k,m}$ is the channel gain for interference link from $SM\#x$ that served by $DC\#k$ to the $DC\#l$ on $channel\#\,m$. Therefore, the maximum acceptable interference on $SM\#n$ served by $DC\#l$ ($I_{max,n}^l$) can be calculated as

$$I_{max,n}^l = \frac{P_s.h_{n,l}^m}{\Gamma_s} - W^m, \quad 1 \le l \le L, and\ 1 \le n \le N \quad (6)$$

Based on (4) and (6), the elements of the reusing candidates' matrix $R$ can be calculated as

$$r_{n,m}^l = \begin{cases} 1, & if\ P_s.h_{n,eNB}^{l,m} \le I_{max,eNB}^m\ and \\ & P_c.h_{m,l}^m \le I_{max,n}^l \\ 0, & oteherwise \end{cases} \quad (7)$$

### 3.2. Phase 2: Optimal Radio Resource Allocation

In the aforementioned part, we have showed how to determine the various reuse candidates for each SM with aimed quality of service constraints for SMs and CUEs. The best reuse partner for every SM once more than one partner CUEs are available is found in this part. The suggested method aims at maximizing the total number of accepted SMs, the optimal RRA problem is modelled as a maximum bipartite matching problem. Fig. 2 illuminates the maximum weight bipartite matching problem in (2), where the set of CUEs and the union of all the reuse candidate of SMs in all DCs are supposed as the two sets of vertices in the bipartite chart. Vertex n in DC l is linked with vertex m by an edge $r_{n,m}^l$, if and only if the $CUE\#m$ is a reuse candidate of $SM\#n$ in $DC\#l$.

In order to select the optimal reuse candidate for every SM with low complexity, the proposed algorithm begins with the SM getting the least appearance (fewest edges) and gets its match with the CUE having fewest edges. Then, we move down the list to the next SM with the following fewest number of edges and so on. Then, the elements of the optimal RRA matrix $A$ are determined using (8).

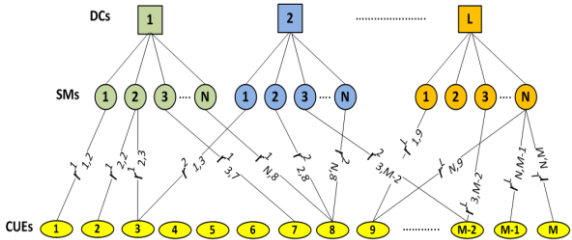

***Fig**. 2. Bipartite graph for SMs and the reuse candidates matching problem.*

$$a_{n,m}^l = \begin{cases} 1, & if\ \sum_{l=1}^{L}\sum_{n=1}^{N} r_{n,m}^l.P_s.h_{n,eNB}^{l,m} \le I_{max,eNB}^m\ and \\ & P_c.h_{m,l}^m + \sum_{\substack{k=1 \\ k \ne l}}^{L}\sum_{x=1}^{N} r_{x,m}^k.P_s.h_{x,l}^{k,m} \le I_{max,n}^l \\ 0, & oteherwise \end{cases} \quad (8)$$

Because we always move to the residual least appearance SM, this process gives better chance for overall matching success without iteration. This also gives certain fairness guarantee for SMs with fewer connections. The details of the proposed RRA algorithm for SMs are shown in Algorithm 1.

---

**Algorithm 1:** RRA for SMs

---
1: $L$: The number of DCs.
2: $M$: The number of CUEs.
3: $N$: The number of SMs per each DC.
4: $\Gamma_c^m$: Minimum acceptable SINR received at eNB for each $CUE\#m$.
5: $\Gamma_s$: Minimum acceptable SINR received at DC for SM.
8: Set $r_{n,m}^l = 0$ and $a_{n,m}^l = 0\ 1 \le m \le M, 1 \le n \le N\ and\ 1 \le l \le L$.
9: **Begin**
10: **Step 1: Find reuse candidates for each SM in each DC**
11:   **For** $m = 1:M$ do
12:     **For** $l = 1:L$ do
13:       **For** $n = 1:N$ do
14:         **if** $P_s.h_{n,eNB}^{l,m} \le I_{max,eNB}^m$ and $P_c.h_{m,l}^m \le I_{max,n}^l$ then
15:           $r_{n,m}^l = 1$
16:         end
17:       end
18:     end
19:   end
20: **Step 2**
21: Find number of times each SM and each CUE was selected
22:   **For** $m = 1:M$ do
23:     $Num_c^m = Sum(r_{1:N,m}^{1:L})$
24:   end
25: Sort the $Num_c^m$ in descending order
26:   **For** $l = 1:L$ do
27:     **For** $n = 1:N$ do
28:       $Num_n^l = Sum\left(r_{n,1:M}^l\right)$
29:     end
30:   end
31: **Step 3**
32: Select the $SM\#n^*$ in $DC\#l^*$ with minimum $Num_{n^*}^{l^*}$ and $Num_{n^*}^{l^*} \ne \infty$.
33: **For** $m = 1:M$ do
34:   **If** $r_{n^*,m}^{l^*} = 1$ then
35:     Choose $CUE\#m$ with minimum $Num_c^m$ for $SM\#n^*$ in $\#l^*$
36:     Set $a_{n^*,m}^{l^*} = 1$
37:     Set $r_{n,m}^{l^*}=0$ for $1 \le n \le N$ except $n=n^*$
38:     Compute the SINR for $CUE\#m$, $SM\#n^*$ and all SMs reusing $channel\#m$, and compare with their own SINR requirements;
39:     **If** all the SINR requirements are satisfied, Then
40:       $Num_{n^*}^{l^*} = \infty$
41:       break
42:     **else**
43:       $a_{n^*,m}^{l^*} = 0$
44:     end
45:   end
46: **end**
47: **Return** to **step 3** until $l = 0$
48: **end**

## 4. Simulation Analysis and Discussion

Simulation results are presented to assess the performance of the suggested RRA algorithm. Firstly, the two metrics (the throughput gain and the access rate) used in the estimation procedure are described. Afterwards, the suggested scheme is estimated by using these two metrics with different network factors. The access rate and the throughput gain are calculated to evaluate the performance of the suggested algorithm. The access rate is the ratio of the number of operated SMs in the D2D reuse mode to the overall number of SMs in the system. The throughput gain is the rise in the total system throughput as a result of sharing of the CUEs UL RRs by SMs. The throughput gain can be calculated as

$$\begin{aligned}TG &= \sum_{m=1}^{M} \log_2\left(1 + \frac{P_c \cdot h_{m,eNB}^m}{W^m + \sum_{k=1}^{L}\sum_{n=1}^{N} a_{n,m}^k \cdot P_s \cdot h_{n,eNB}^{k,m}}\right) \\ &+ \sum_{l=1}^{L}\sum_{n=1}^{N}\sum_{m=1}^{M} a_{n,m}^l \cdot \log_2\left(1 \right.\\ &+ \left. \frac{P_s \cdot h_{n,l}^m}{P_c \cdot h_{m,l}^m + \sum_{\substack{k=1\\k\neq l}}^{L}\sum_{x=1}^{N} a_{x,m}^k \cdot P_s \cdot h_{x,l}^{k,m} + W^m}\right) \\ &- \sum_{m=1}^{M} \log_2\left(1 + \frac{P_c \cdot h_{m,eNB}^m}{W^m}\right)\end{aligned} \quad (9)$$

We supposed that CUEs are distributed uniformly inside the cell and each group of SMs are assumed to be uniformly distributed around its serving DC in accordance with the maximum definite distance between SMs and its serving DC. In Table I, Simulation parameters are mentioned.

Fig. 3 illustrates the distribution of SMs, DCs, and CUEs in the circular cellular cell with radius of 0.5 km. The eNB is situated at the origin of the cell and the locations of DCs are decided as a chosen point randomly on the circumstance of a circle with a radius equal to the identified distance between DC and eNB.

Fig. 4 evaluates the throughput gain and access rate of the network at different values for $Md_{SD}$. As can be illustrated from Fig. 4(a) and Fig. 4(b), the access rate and consequently the throughput gain are decreased when $Md_{SD}$ is increased. This is because the received signals strength from SMs at its serving DC will drop due to the increasing distance of SM/DC links. Consequently, the interference from CUEs communications at the DC will be more affective which declines the received SINRs of SMs. This decreases the possibility of reusing the RRs by SMs that not attaining its quality of service requirements that will result in a lessening in the throughput gain and access rate.

As shown in Fig.4 (a), the proposed network can admit more than 90% of the existed number of SMs to work in a reuse mode, whereas attaining the mandatory quality of service requirements for CUEs and SMs when $Md_{SD}$ does not exceed 50m. This access rate declined to 70% once $Md_{SD}$ reaches 75m. In comparison with the other short-range wireless communication technologies which have nearly a comparable communication distance, the suggested network can provision the mandatory SMs communication by controlling the interference and getting automatic pairing and security.

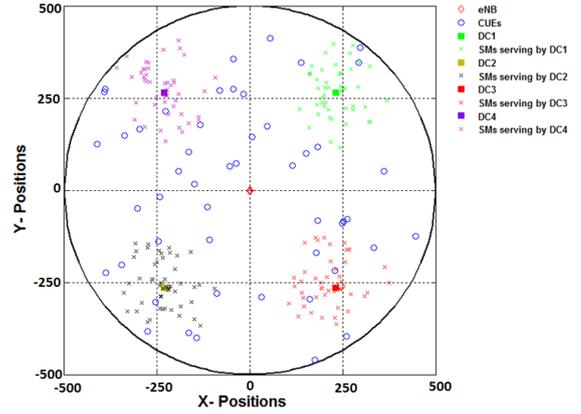

*Fig. 3.* Snapshot for SMs, DCs, and CUEs distribution in a single cell when $M = 50, L = 4, N = 50$ and maximum distance between SM and its serving DC ($Md_{SD}$)=50m.

**TABLE I** Simulation Parameters.

| Parameter | Value |
|---|---|
| Number of channels (RBs) | 50 |
| Number of CUEs (M) | 50 |
| UL System Bandwidth | 10MHz |
| Number of DCs in the network (L) | 4 |
| Number of SMs per each DC (N) | 10%,20%,…..,200% of CUEs |
| Bandwidth of each channel (RB) | 180KHz |
| Noise power spectral density ($N_0$) | $-174$ dBm/Hz |
| Cell radius | 500 m |
| SINR Requirement for CUEs | Uniform distributed from 0 to 25 dB |
| SINR Requirement for SMs | 5dB |
| Maximum distance between SM and serving DC ($Md_{SD}$) | 50m, 75m and 100m |
| Maximum transmit power of SMs ($P_s^{max}$) | 24dBm |
| Path loss model for Cellular links | $128.1 + 37.6 log10\ d\ (dB)$, d is the distance between transmitter and receiver in kilometres |
| Path loss model between SM and serving DC | $148 + 40 log10\ d\ (dB)$, d is the distance between transmitter and receiver in kilometres |
| Maximum transmit power of CUEs ($P_c^{max}$) | 24dBm |

Since CUEs have various applications that need various quality of service levels, the performance of the network is estimated at various SINR requirements for CUEs. In Fig. 5, higher levels of throughput gain and access rate can be acquired when CUEs with low SINR constraints. Since lower levels of SINR requirements implies higher interference limitations are permissible for CUEs and subsequently larger numbers of SMs reuse candidates are acceptable for each CUE. This causes increasing for the possibility of SMs to work in the D2D reuse mode and therefore, higher throughput gain and access rate are gotten.

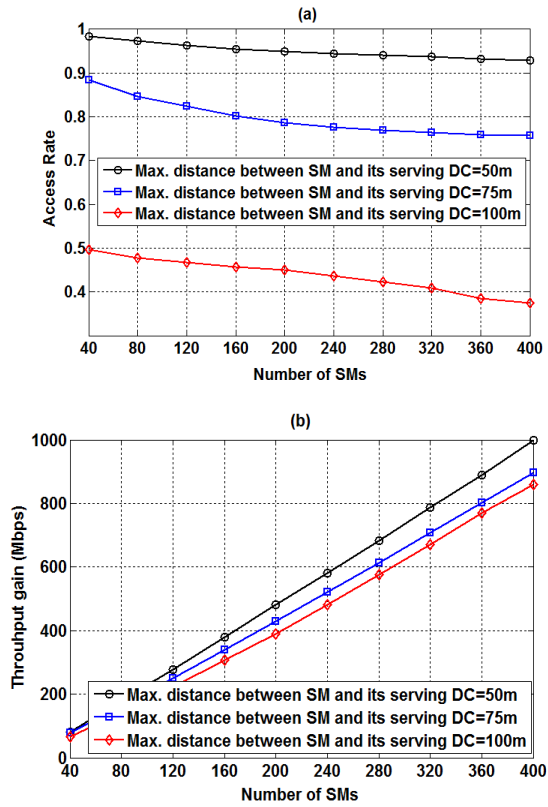

*Fig. 4. Access rate and throughput gain vs. Number of SMs.*

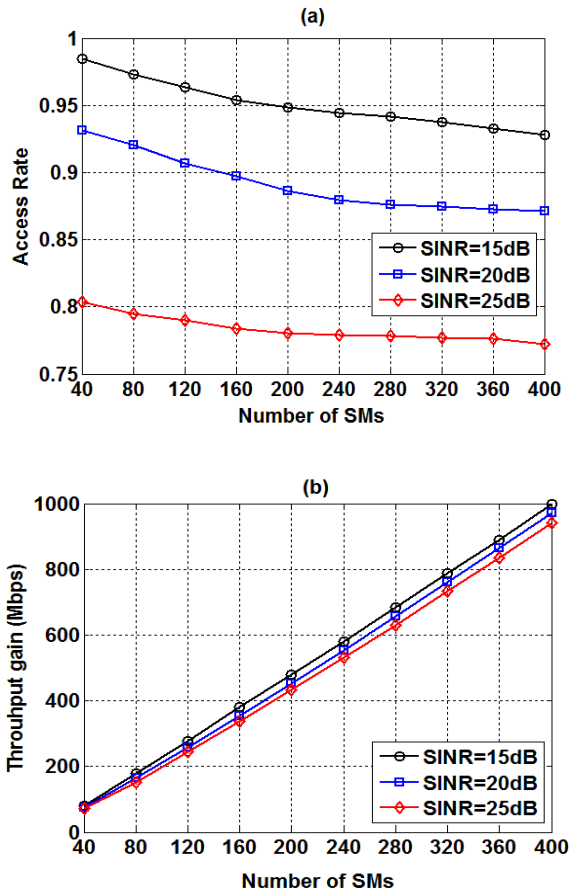

*Fig. 5. Access rate and throughput gain vs. Number of SMs when $Md_{SD} = 50m$.*

## 5. Conclusion

A novel approach is suggested to allow SMs to communicate with its serving data concentrator using D2D communication technology underlaying LTE systems in this paper. The suggested algorithm is dependent on reusing the UL RRs allocated for CUEs in the cellular networks to provision the mandatory communication between a set of SMs and its serving DC. To control the interference between SMs and CUEs, and achieve the mandatory quality of service requirements for CUEs and SMs; two phases based UL RRA approach for SMs with quality of service provisioning is suggested. The case of more than one SM sharing the same channel is considered. The group of permissible cellular reuse candidates for each SM is decided in the first phase. In other words, the interference between SMs and CUEs is deemed and coordinated to ensure the mandatory quality of service requests for SMs and CUEs. The optimal RRA problem is modelled by using bipartite graph to find the optimal reuse partner for each SM aiming at maximizing the overall number of SMs which can be accepted in the second phase. The suggested algorithm can provide the required connectivity between SMs and its serving DC underlaying cellular systems with efficient performance in terms of the throughput gain and the network access rate as can be seen in the simulation results. The proposed network can admit more than 90% of the existed number of SMs to work in a reuse mode with guaranteed quality of service constraints for both CUEs and SMs. Furthermore, the benefits of organized interference, security and fixed automatic pairing of the suggested approach make it a promising solution in comparison with the other wireless communication technologies for SMs communications.